\documentclass[10pt]{article}
\usepackage{latexsym}
\usepackage{amssymb}
\usepackage{amsmath}
\usepackage{amscd}
\usepackage{amsthm}
\usepackage[left=2cm,top=2.5cm,right=2.5cm,bottom=1.5cm]{geometry}

\usepackage[dvips]{graphicx}
\usepackage{hyperref}
\begin{document}
\begin{center}
\large{\bf {Generation of Bianchi Type V String Cosmological Models with Bulk Viscosity}} \\
\vspace{10mm}
\normalsize{Anil Kumar Yadav}\\
\vspace{4mm}
\normalsize{Department of Physics, Anand Engineering
College, Keetham, Agra-282 007, India} \\
\vspace{2mm}
\normalsize{E-mail: abanilyadav@yahoo.co.in}\\
%\vspace{5mm}
%\normalsize{}
%\vspace{5mm}
%\normalsize{}
\end{center}
\vspace{8mm}
%\date{}
%\maketitle
\begin{abstract}
Bianchi type V string cosmological models with bulk viscosity for massive string are investigated.
The bulk viscosity is assumed to vary with time in such a manner that it is related to simple power
function of the energy density. Using generation technique (Camci~et~al., 2001),
it is shown that Einstein's field equations are solvable for
any arbitrary cosmic scale function. Solution for particular form of cosmic scale functions are also obtained.
It is found that solutions based on generation technique are relevant to the observational results. Some physical 
and geometrical aspect of the models are also discussed.\\
\end{abstract}
\section{Introduction}
The string theory plays a significant role in the study of physical situation at the very early
stages of the formation of the universe. It is generally assumed that after the big bang, 
the universe may have undergone a series of phase transitions as its temperature lowered down below some 
critical temperature as predicted by grand unified theories \cite{ref1}$-$\cite{ref6}. At the very early stages of evolution  
of the universe, it is believed that during phase transition the symmetry of the universe is broken spontaneously. 
It can give rise to topologically stable defects such as domain walls, strings and monopoles \cite{ref1}. 
In all these three cosmological structures, only cosmic strings have excited the most interesting
consequence \cite{ref7} because they are believed to give rise to density perturbations which lead 
to formation of galaxies \cite{ref4,ref8}. These cosmic strings can be closed like loops or open 
like a hair which move through time and trace out a tube 
or a sheet, according to whether it is closed or open. The string is free to vibrate and its 
different vibrational modes present different types of particles carrying the force of gravitation. 
This is why much interesting to study the gravitational effect that arises from strings by using Einstein's 
field equations. The general relativistic treatment of strings has been initially given by Letelier \cite{ref9,ref10} and 
Stachel \cite{ref11}. Letelier \cite{ref9} obtained the general solution of Einstein's field equations for a cloud of
 strings with spherical,plane and a particular case of cylindrical symmetry. Letelier \cite{ref10} also obtained massive string
cosmological models in Bianchi type-I and Kantowski-Sachs space-times. Benerjee
et al. \cite{ref12} have investigated an axially symmetric Bianchi type I string dust
cosmological model in presence and absence of magnetic field. Exact solutions of
string cosmology for Bianchi type II, $VI_{0}$, VIII and IX space-times have
been studied by Krori et al. \cite{ref13} and Wang \cite{ref14,ref15}. Bali and Upadhaya \cite{ref16} 
have presented LRS Bianchi type I string dust magnetized  cosmological models. Singh  and Singh \cite{ref17} investigated string cosmological 
models with magnetic field in the context of space-time with $G_{3}$ symmetry. Singh \cite{ref18,ref19}, has studied 
string cosmology with electromagnetic fields in Bianchi type II, VIII and IX space-times.

Bianchi V universes are the natural generalization of FRW models with negative curvature.These open models 
are favored by the available evidences for low density universes (Gott et al \cite{ref20}).Bianchi type V 
cosmological model 
where matter moves orthogonally to the hypersurface of homogeneity, has been studied by Heckmann and 
Schucking \cite{ref21}.
Exact tilted solutions for the Bianchi type V space-time are obtained by 
Hawkings \cite{ref22}, Grishchuk et al. \cite{ref23}. Ftaclas
and Cohen \cite{ref24} have investigated LRS Bianchi type V universes containing stiff matter with electromagnetic field.
Lorentz \cite{ref25} has investigated LRS Bianchi type V tilted models with stiff fluid and electromagnetic field. 
Pradhan et al \cite{ref26} have investigated The generation of Bianchi type V cosmological models with varying $\Lambda$ 
term. Yadav et al. \cite{ref27,ref28} have investigated bulk viscous string cosmological models in 
different space-times. Bali and Anjali \cite{ref29}, Bali \cite{ref30} have obtained 
Bianchi type-I, and type V string cosmological models in general 
relativity. The string cosmological models with a magnetic field are discussed by Tikekar and Patel \cite{ref31},
Patel and Maharaj \cite{ref32}. Ram and Singh \cite{ref33} obtained some new exact solution of string 
cosmology with and without a source free magnetic field for Bianchi type I space-time in the different basic form 
considered by Carminati and McIntosh \cite{ref34}. Yavuz et al. \cite{ref35} have examined 
charged strange quark matter attached to the string cloud in the spherical symmetric 
space-time admitting one-parameter group of conformal motion. Kaluza-Klein 
cosmological solutions are obtained by Yilmaz \cite{ref36} for quark matter attached to the 
string cloud in the context of general relativity.

The distribution of matter can be satisfactorily described by a perfect fluid
due to the large scale distribution of galaxies in our universe. However, observed
physical phenomena such as the large entropy per baryon and the remarkable
degree of isotropy of the cosmic microwave background radiation, suggest anal-
ysis of dissipative effects in cosmology. Furthermore, there are several processes
which are expected to give rise to viscous effects. These are the decoupling of
neutrinos during the radiation era and the decoupling of radiation and matter
during the recombination era. Bulk viscosity is associated with the GUT phase
transition and string creation. Misner \cite{ref37} has studied the effect of viscosity on
the evolution of cosmological models. The role of viscosity in cosmology has
been investigated by Weinberg \cite{ref38}. Nightingale \cite{ref39}, Heller and Klimek \cite{ref40}
have obtained a viscous universes without initial singularity. The model stud-
ied by Murphy \cite{ref41} possessed an interesting feature in which big bang type of
singularity of infinite space-time curvature does not occur to be a finite past.
However, the relationship assumed by Murphy between the viscosity coefficient
and the matter density is not acceptable at large density. Thus, we should con-
sider the presence of material distribution other than a perfect fluid to obtain a
realistic cosmological models (see Gr$\o{}$n \cite{ref42} for a review on cosmological models
with bulk viscosity). The effect of bulk viscosity on the cosmological evolution
has been investigated by a number of authors in the framework of general theory
of relativity.

%%%%%%%%%%%%%%%%%%%%%%%%%%%%%%%%%%%%%%%%%%%%%%%%%%%%%%%%%%%%%%%%%%%%%%%%%%%%%%%%%%%%%%%%%%%%%%%%%%%%%%%%%%%%%
%%%%%%%%%%%%%%%%%%%%%%Field Equation%%%%%%%%%%%%%%%%%%%%%%%%%%%%%%%%%%%%%%%%%%%%%%%%%%%%%%%%%%%%%%%%%%%%%%%%%
\section{Field Equations}
we consider the Bianchi type V metric of the form
\begin{equation}
\label{eq1} ds^2=dt^2-A^2(t)dx^2-e^{2\alpha x} [B^2(t)dy^2+C^2(t)dz^2]
\end{equation}
where $\alpha$ is a constant.\\ 
The energy momentum tensor for cloud of string dust with bulk viscous fluid of string given by
Landau and Lifshitz (1963) and Letelier (1979)
\begin{equation}
\label{eq2} T^{j}_{\;i} =\rho v_{i}v^{j} - \lambda x_{i}x^{j} - \xi v;^{l}_{l}\left(g^{j}_{\;i}+v_{i}v^{j}\right)
\end{equation}
where $v_{i}$ and $x_{i}$ satisfy the condition
\begin{equation}
\label{eq3} v^{i}v_{i} = -x^{i}x_{i} = 1,  v^{i}x_{i} = 0.
\end{equation}
Here $\rho$ is the proper energy density of the cloud of string with
particle attached to them.$\lambda$ is the string tension density,
$v^{i}$, the four velocity of the particles and $x^{i}$, the
unit space vector representing the direction of strings.If the
particle density of the configuration is denoted by $\rho_{p}$ then
we have
\begin{equation}
\label{eq4} \rho = \rho_{p} +\lambda
\end{equation}
For the energy momentum tensor (\ref{eq2}) and Bianchi type V metric (\ref{eq1}),
Einstein's field equations
\begin{equation}
\label{eq5} R^{j}_{\;i}-\frac{1}{2}Rg^{j}_{\;i} = -8\pi T^{j}_{\;i}
\end{equation}
yield the following five independent equations
\vspace{3mm}
\begin{equation}
\label{eq6} \frac{\ddot{A}}{A}+\frac{\ddot{B}}{B}+\frac{\dot{A}\dot{B}}{AB}-\frac{\alpha^2}{A^2}=-8\pi\xi\theta
\end{equation}
\begin{equation}
\label{eq7} \frac{\ddot{A}}{A}+\frac{\ddot{C}}{C}+\frac{\dot{A}\dot{C}}{AC}-\frac{\alpha^2}{A^2}=-8\pi\xi\theta
\end{equation}
\begin{equation}
\label{eq8} \frac{\ddot{B}}{B}+\frac{\ddot{C}}{C}+\frac{\dot{B}\dot{C}}{BC}-\frac{\alpha^2}{A^2}=-8\pi
\left(\lambda+\xi\theta\right)
\end{equation}
\begin{equation}
\label{eq9} \frac{\dot{A}\dot{B}}{AB}+\frac{\dot{A}\dot{C}}{AC}+\frac{\dot{B}\dot{C}}{BC}-\frac{3\alpha^2}{A^2}=-8\pi
\rho
\end{equation}
\begin{equation}
\label{eq10} \frac{2\dot{A}}{A}-\frac{\dot{B}}{B}-\frac{\dot{C}}{C}=0
\end{equation}
Here and what follows the dots overhead the symbol A, B, C denotes
differentiation with respect to t.\\
The physical quantities expansion scalar $\theta$ and shear scalar $\sigma^2$ have the following expressions
\begin{equation}
\label{eq11} \theta = \frac{\dot{A}}{A}+\frac{\dot{B}}{B}+\frac{\dot{C}}{C}
\end{equation}
\begin{equation}
\label{eq12} \sigma^2 =\frac{1}{2}\sigma_{ij}\sigma^{ij}=\frac{1}{3}\left[\theta^2-\frac{\dot{A}\dot{B}}{AB} 
-\frac{\dot{A}\dot{C}}{AC}-\frac{\dot{B}\dot{C}}{BC}\right]
\end{equation}
 
Integrating eqs.(\ref{eq10}) and absorbing the integrating constant into B or C, we obtain
\begin{equation}
\label{eq13} A^2=BC
\end{equation}
without loss of any generality. Now subtracting equation $(7)$ from $(6)$, we obtain
\begin{equation}
\label{eq14} 2\frac{\ddot{B}}{B}+\left(\frac{\dot{B}}{B}\right
)^2 = 2\frac{\ddot{C}}{C}+\left(\frac{\dot{C}}{C}\right)^{2}
\end{equation}
which on integration yields
\begin{equation}
\label{eq15} \frac{\dot{B}}{B}-\frac{\dot{C}}{C} = \frac{k}{(BC)^{\frac{3}{2}}}
\end{equation}
where k is the constant of integration. Hence for the metric function B or C from the above first order differential
eqs.(\ref{eq15}), some scale transformations permit us to obtain new metric function B or C.\\
~~~ Firstly, under the scale transformation $dt=B^{\frac{1}{2}}dT$, equation $(15)$ takes the form
\begin{equation}
\label{eq16} CB_{T}-BC_{T} = kC^{-\frac{1}{2}}
\end{equation}
where the subscript represents derivative with respect to $T$. Considering eqs.(\ref{eq16}) as a linear
differential equation for B, where C is an arbitrary function, we obtain
\begin{equation}
\label{eq17} (i)\;\;\;\;\;\;\;\;B = k_{1}C+kC \int{\frac{dT}{C^{\frac{5}{2}}}},
\end{equation}
where $k_{1}$ is the the constant of integration. Similarly, using the transformation $dt=B^\frac{3}{2}d\bar{T}$,
$dt=C^\frac{1}{2}d\tilde{t}$ and $dt=C^\frac{3}{2}d\tau$ in equation (\ref{eq13})
after some algebra we obtain respectively.
\begin{equation}
\label{eq18} (ii)\;\;\;\;\;\;\;\;\;\; B(\bar{T})=k_{2}Ce^{\left(k\int\frac{d\bar{T}}{C^\frac{3}{2}}\right)},
\end{equation}
\begin{equation}
\label{eq19} (iii)\;\;\;\;\;\;\;\;\;\;\; C(\tilde{t})=k_{3}B-kB\int\frac{d\tilde{t}}{B^\frac{5}{2}},
\end{equation}
\begin{equation}
\label{eq20}(iv)\;\;\;\;\;\;\;\;\;\;\;\; C(\tau)=k_{4}Be^{\left(k\int\frac{d\tau}{B^\frac{3}{2}}\right)},
\end{equation}
where $k_{2}$,$k_{3}$ and $k_{4}$ are constant of integration. Thus choosing any given function B or C in 
cases (i), (ii), (iii) and (iv), one can obtain B or C.\\
%%%%%%%%%%%%%%%%%%%%%%%%%%%%%%%%%%%%%%%%%%%%%%%%%%%%%%%%%%%%%%%%%%%%%%%%%%%%%%%%%%%%%%%%%%%%%%%%%%%%%%%%%%%%%%%%%%
%%%%%%%%%%%%%%%%%%% Generation technique for solution %%%%%%%%%%%%%%%%%%%%%%%%%%%%%%%%%%%%%%%%%%%%%%%%%%%%%%%%%%
\section{Generation technique for solution}
We consider the following four cases
%%%%%%%%%%%%%%%%%%%%%%%%%%%%%%%%%%%%%%%%%%%%%Subsection3.1%%%%%%%%%%%%%%%%%%%%%%%%%%%%%%%%%%%%%%%%%%%%%%%%%%%%%%%%%%
%%%%%%%%%%%%%%%%%%%%%%%%%%%%%%%%%%%%%%%%%%%%%%%%%%%%%%%%%%%%%%%%%%%%%%%%%%%%%%%%%%%%%%%%%%%%
\subsection{Case (i):~$ C=T^n $~~(n~is~a~real~number~satisfying~$ n\neq\frac{2}{5}$)}
In this case equation (\ref{eq15}) gives
\begin{equation}
\label{eq21} B=k_{1}T^n+\frac{2k}{2-5n}T^{{1}-\frac{3n}{2}}
\end{equation}
and then equation (\ref{eq11}), we obtain
\begin{equation}
\label{eq22} A^2=k_{1}T^{2n}+\frac{2k}{2-5n}T^{1-\frac{n}{2}}
\end{equation}
Hence metric (\ref{eq1}) reduces to the following form
\begin{equation}
\label{eq23} ds^2=(k_{1}T^n+2lT^{l_{1}})[dT^2-T^ndx^2]-e^{2\alpha x}\left[\left(k_{1}T^n+2lT^{l_{1}}\right)^2dy^2+T^{2n}dz^2\right],
\end{equation}
where $l=\frac{k}{2-5n}$ and $l_{1}=1-\frac{3n}{2}$\\
In this case the physical parameters, i.e. the string tension density $(\lambda)$, the energy density $(\rho)$, 
the particle density $(\rho_{p})$ and the kinematical parameters, i.e. the scalar of expansion $(\theta)$,
shear scalar $(\sigma)$ and the proper volume $(V^3)$ for model $(23)$ are given by\\
\[
8\pi\lambda=\left[-2k_{1}^2n(n-1)T^{2n-2}-k_{1}ln(10-13n){T^{-(l_{1}+2n)}}-\frac{1}{2}l^2(4+4n-11n^2){T}^{-3n}\right]
\psi_{1}^{-3}
\]
\begin{equation}
\label{eq24} + \alpha^2{T}^{-n}\psi_{1}^{-1} + 24\pi \xi\left[k_{1}nT^{n-1}+
\frac{1}{2}l(2-n)T^{\frac{-3n}{2}}\right]\psi_{1}^{-\frac{3}{2}},
\end{equation}
\[
8\pi\rho=\left[3k_{1}^2n^2T^{2n-2}+3k_{1}ln(2-n)T^{-(l_{1}+2n)}+\frac{1}{2}l^2(4+4n-11n^2)T^{-3n}\right]\psi_{1}^{-3}
\]
\begin{equation}
\label{eq25} -3\alpha^2T^{-n}\psi_{1}^{-1},
\end{equation}
\[
8\pi\rho_{p}=\left[n(5n-2)k_{1}^2T^{2n-2}+16k_{1}ln(1-n)T^{-(l_{1}+2n)}+l^2(4+4n-11n^2)T^{-3n}\right]\psi_{1}^{-3}
\]
\begin{equation}
\label{eq26} - 4\alpha^2T^n\psi_{1}^{-1} - 24\pi \xi\left[k_{1}nT^{n-1}+
\frac{1}{2}l(2-n)T^{\frac{-3n}{2}}\right]\psi_{1}^{-\frac{3}{2}}
\end{equation}
\begin{figure}
\begin{center}
\includegraphics[width=4.0in]{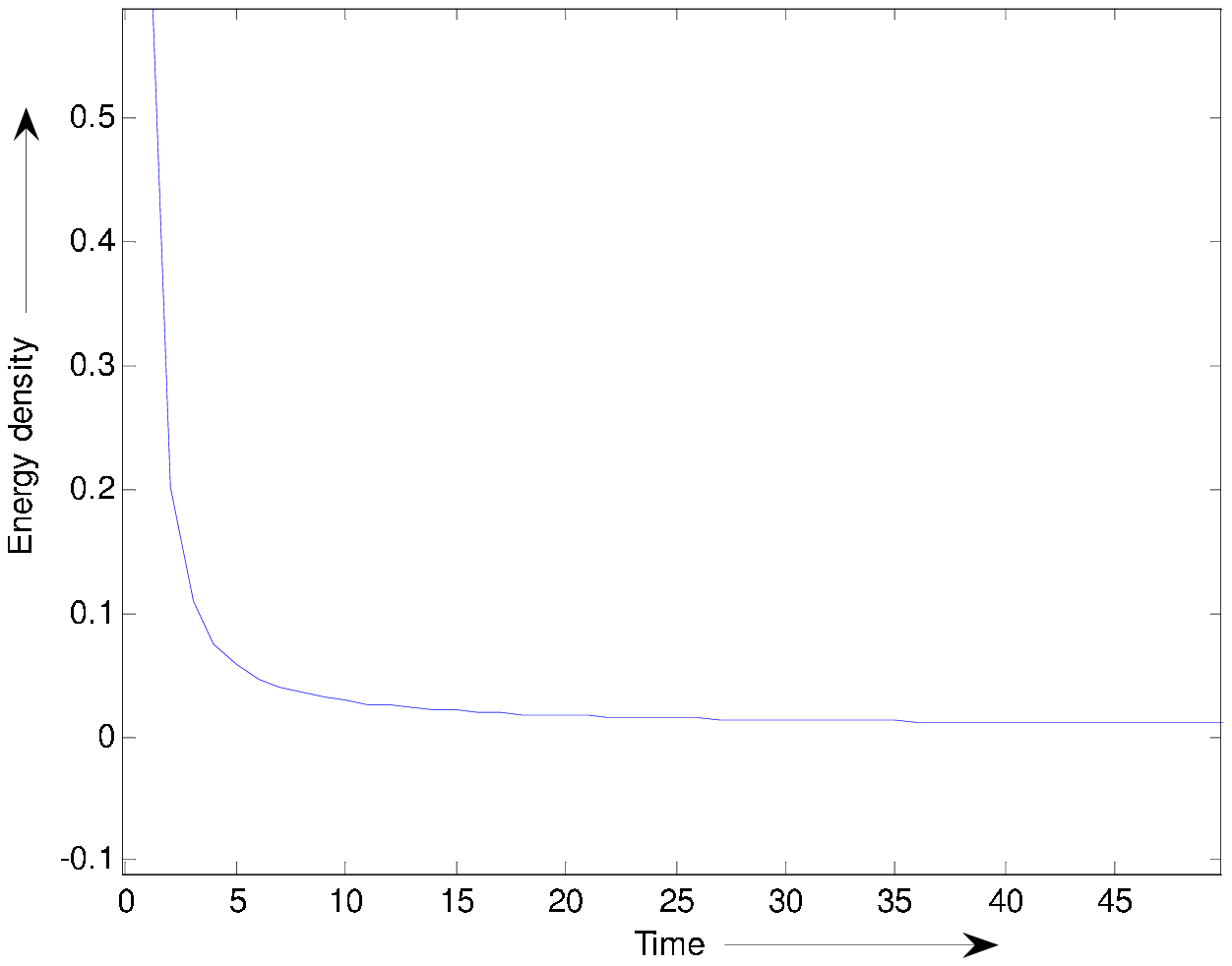} 
\caption{The plot of energy density $\rho$ vs. time T with parameters $ n = 0.5, k = 1, \alpha = 1 
 k_{1} = 1, l_{1} = 0.25 $}
\label{fg:anil22F1.eps}
\end{center}
\end{figure}

Where $\psi_{1} = \left(k_{1}T^n+2lT^{l_{1}}\right)$.\\
This general solution has a rich structure and admits many numbers of solutions by suitable choice of $ \xi(t) $. 
Here the choice of $\xi(t)$ is quite arbitrary but since we look for physically viable model of universe thus for 
specification of $\xi$, in most of the investigations, involving bulk viscosity, it is assumed to be simple power 
function of energy density (Pavon \cite{ref43} , Maartens \cite{ref44} and Zimdahl \cite{ref45}).
\begin{equation}
\label{eq27} \xi(t) = \xi_{0}\rho^w
\end{equation}
Where $\xi_{0}$ and $w$ are constant. Now we consider two cases depending on two values of w, namely $0$ and $1$.
%%%%%%%%%%%%%%%%%%%%%%%%%%%%%%%%%%%%%%%%%%%%%%%%%%%%%%%%%%%%%%%%%%%%%%%%%%%%%%%%%%%%%%%%%%%%%%%%%%%%%%%%%%%
%%%%%%%%%%%%%%%%%%%%%%%%%%%%%%%%Subsubsection 3.1.1 %%%%%%%%%%%%%%%%%%%%%%%%%%%%%%%%%%%%%%%%%%%%%%%%%%%%%%%
\subsubsection{Case I:~ Solution for $\xi = \xi_{0}$} 
For $ w = 0 $, eqs.(\ref{eq27}) reduces $\xi=\xi_{0}$.Thus eqs.(\ref{eq24}) and eqs.(\ref{eq26}) reduces to 
\[
8\pi\lambda=\left[-2k_{1}^2n(n-1)T^{2n-2}-k_{1}ln(10-13n){T^{-(l_{1}+2n)}}-\frac{1}{2}l^2(4+4n-11n^2){T}^{-3n}\right]
\psi_{1}^{-3}
\]
\begin{equation}
\label{eq28} + \alpha^2{T}^{-n}\psi_{1}^{-1} + 24\pi \xi_{0}\left[k_{1}nT^{n-1}+
\frac{1}{2}l(2-n)T^{\frac{-3n}{2}}\right]\psi_{1}^{-\frac{3}{2}},
\end{equation}
 \[
8\pi\rho_{p}=\left[n(5n-2)k_{1}^2T^{2n-2}+16k_{1}ln(1-n)T^{-(l_{1}+2n)}+l^2(4+4n-11n^2)T^{-3n}\right]\psi_{1}^{-3}
\]
\begin{equation}
\label{eq29} - 4\alpha^2T^n\psi_{1}^{-1} - 24\pi \xi_{0}\left[k_{1}nT^{n-1}+
\frac{1}{2}l(2-n)T^{\frac{-3n}{2}}\right]\psi_{1}^{-\frac{3}{2}}
\end{equation}
%%%%%%%%%%%%%%%%%%%%%%%%%%%%%%%%%%%%%%%%%%%%%%%%%%%%%%%%%%%%%%%%%%%%%%%%%%%%%%%%%%%%%%%%%%%%%%%%%%%%%%%%%%%%%%%%%%%
%%%%%%%%%%%%%%%%%%%%%%%%%%%%%%%%%%% Subsubsection 3.1.2 %%%%%%%%%%%%%%%%%%%%%%%%%%%%%%%%%%%%%%%%%%%%%%%%%%%%%%%%%%%
\subsubsection{Case II:~ Solution for $\xi = \xi_{0}\rho$}
For $w=1$, eqs.(\ref{eq27}) reduces $\xi=\xi_{0}\rho$, with use of eqs.(\ref{eq25}), eqs.(\ref{eq24}) and 
eqs.(\ref{eq26}) reduces to
\[
8\pi\lambda=\left[-2k_{1}^2n(n-1)T^{2n-2}-k_{1}ln(10-13n){T^{-(l_{1}+2n)}}-\frac{1}{2}l^2(4+4n-11n^2){T}^{-3n}\right]
\psi_{1}^{-3}
\]
\[
 + \alpha^2{T}^{-n}\psi_{1}^{-1}-9\xi_{0}\alpha^2T^{-n}\left(k_{1}nT^{n-1}+
\frac{1}{2}l(2-n)T^{\frac{-3n}{2}}\right)\psi_{1}^{-\frac{5}{2}} + 
\]
\[
 3\xi_{0}\left(3k_{1}^2n^2T^{2n-2}+3k_{1}ln(2-n)T^{-(l_{1}+2n)}+
\frac{1}{2}l^2(4+4n-11n^2)T^{-3n}\right)\psi_{1}^{-3}
\]

\begin{equation}
\label{eq30} \left(k_{1}nT^{n-1}+
\frac{1}{2}l(2-n)T^{\frac{-3n}{2}}\right)\psi_{1}^{-\frac{3}{2}}
\end{equation}
 \[
8\pi\rho_{p}=\left[n(5n-2)k_{1}^2T^{2n-2}+16k_{1}ln(1-n)T^{-(l_{1}+2n)}+l^2(4+4n-11n^2)T^{-3n}\right]\psi_{1}^{-3}
\]
\[
 - 4\alpha^2T^n\psi_{1}^{-1}+9\xi_{0}\alpha^2T^{-n}\left(k_{1}nT^{n-1}+
\frac{1}{2}l(2-n)T^{\frac{-3n}{2}}\right)\psi_{1}^{-\frac{5}{2}} - 
\]
\[
 3\xi_{0}\left(3k_{1}^2n^2T^{2n-2}+3k_{1}ln(2-n)T^{-(l_{1}+2n)}+\frac{1}{2}l^2(4+4n-11n^2)T^{-3n}\right)
\psi_{1}^{-3}
\]
\begin{equation}
\label{eq31} \left(k_{1}nT^{n-1}+
\frac{1}{2}l(2-n)T^{\frac{-3n}{2}}\right)\psi_{1}^{-\frac{3}{2}} 
\end{equation}
From Eqs.(\ref{eq25}), we note that $\rho(t)$ is a decreasing function of time
and $\rho > 0$ for all times. This behaviour is clearly depicted in Fig. 1
as a representative case with appropriate choice of constants of integration and
other physical parameters using reasonably well known situations. Figures 1
show this physical behaviours of energy density as a decreasing functions of time.
Also it is observed that bulk viscosity affect the string tension density $\lambda$ and particle density $\rho_{p}$.
\begin{equation}
\label{eq32} \theta=3\left[k_{1}nT^{n-1}+\frac{1}{2}l(2-n)T^{\frac{-3n}{2}}\right]\psi_{1}^{-\frac{3}{2}}
\end{equation}
\begin{equation}
\label{eq33} \sigma=\frac{1}{2}kT^{-\frac{3n}{2}}\psi_{1}^{-\frac{3}{2}}~~~~~~~~~~~~~~~~~~~~~~~~~~~~~~
\end{equation}
\begin{equation}
\label{eq34} V^3=\left(k_{1}T^{2n}+2lT^{n+l_{1}}\right)^\frac{3}{2}e^{2\alpha x}~~~~~~~~~~~~~~~~~~~~~~~~~~~~~~
\end{equation}
Equations (\ref{eq32}) and (\ref{eq33}) leads to
\begin{equation}
\label{eq35} \frac{\sigma}{\theta}=\frac{1}{6}k\left[k_{1}nT^{n-l_{1}} +\frac{1}{2}l(2-n)\right]^{-1}~~~~~~~~~~~~~~~~~~~~~~~
\end{equation}
The energy condition $ \rho\geq0 $ and $ \rho_{p}\geq0 $ satisfy for model (\ref{eq23}). The condition $\rho\geq0 $ and
$ \rho_{p}\geq0 $ leads to\\
\[
\left[3k_{1}^{2}n^{2}T^{3n-2}+3k_{1}ln(2-n)T^{-(l_{1}+n)}+\frac{1}{2}l^2
\left(4+4n-11n^2\right)T^{-2n}\right]
\]
\begin{equation}
\label{eq36} \psi_{1}^{-2}\geq{3\alpha}^2
\end{equation}
\[
 \left[n(2-5n)k_{1}^{2}T^{n-2}+16k_{1}ln(n-1)T^{-(l_{1}+3n)}+l^2\left(11n^2-4n-4\right)T^{-4n}\right]
\]
\begin{equation}
\label{eq37} \psi_{1}^{-2} -24\pi\xi\left(k_{1}nT^{-1}+\frac{1}{2}l(2-n)T^{-\frac{5n}{2}}\right)\psi^{-\frac{1}{2}}
\geq{4\alpha}^2
\end{equation}
respectively.\\
We observe that the string tension density $\lambda\geq0$, leads to
\[
\left[-2k_{1}^{2}n(n-1)T^{3n-2}-k_{1}ln(10-13n)T^{-(l_{1}+n)}-\frac{1}{2}l^2
\left(4+4n-11n^2\right)T^{-2n}\right]\psi_{1}^{-2}
\]
\begin{equation}
\label{eq38}+ 24\pi \xi\left[k_{1}nT^{2n-1}+\frac{1}{2}l(2-n)T^{\frac{-n}{2}}\right]\psi_{1}^{-\frac{1}{2}}
\geq{-\alpha}^2
\end{equation}
Generally the model$(23)$ are expanding, shearing and approaches to isotropy at late time. For $k=0$, 
the model becomes shear free. We observe that as $T\rightarrow\infty$, $V^3\rightarrow\infty$ and $\rho\rightarrow0$
hence volume increases when T increases and the proper energy density of the cloud of string with particle 
attached to them is the decreasing function of time.\\ 
%%%%%%%%%%%%%%%%%%%%%%%%%%%%%%%%%%%%%%%%%%%%%%%%%%%%%%%%%%%%%%%%%%%%%%%%%%%%%%%%%%%%%%%%%%%%%%%%%%%%%%%%%%%%%%%%%%%%%%%%%%%%
%%%%%%%%%%%%%%%%%%%%%%%%%%%%%%%%%%%%%%%%%%%%%Subsection3.2%%%%%%%%%%%%%%%%%%%%%%%%%%%%%%%%%%%%%%%%%%%%%%%%%%%%%%%%%%%%%%%%%
\subsection{case(ii): $ C={\bar{T}}^{n} $~~(n is a real number satisfying  $n\neq\frac{2}{3}$)}
In this case equation (\ref{eq18}) gives 
\begin{equation}
\label{eq39} B=k_{2}{\bar{T}}^{n}e^{M{\bar{T}}^{l_{1}}}
\end{equation}
and from equation (\ref{eq13}), we obtain
\begin{equation}
\label{eq40} A^{2}=k_{2}{\bar{T}}^{2n}e^{M{\bar{T}}^{l_{1}}}
\end{equation}
where $M=\frac{k}{l_{1}}$.
Hence the metric (\ref{eq1}) reduces to the form
\begin{equation}
\label{eq41} ds^2={\bar{T}}^{\frac{4(1-l_{1})}{3}}\left[{\bar{T}}^{\frac{2(1-l_{1})}{3}}e^{3M{\bar{T}}^{l_{1}}}d{\bar{T}^2}
-e^{M{\bar{T}}^{l_{1}}}dx^2-e^{2\alpha x}\left(e^{2M{\bar{T}}^{l_{1}}}dy^2+dz^2\right)\right],
\end{equation}
where the constant $k_{2}$ is taken, without loss of generality, equal to $1$.
In this case the physical parameters, i.e. the string tension density $(\lambda)$, the energy density $(\rho)$, 
the particle density $(\rho_{p})$ and the kinematical parameters, i.e. the scalar of expansion $(\theta)$,
shear scalar $(\sigma)$ and the proper volume $(V^3)$ for model (\ref{eq41}) are given by\\
\[
 8\pi\lambda=2n{\bar{T}}^{2(l_{1}-2)}+3nk{\bar{T}}^{3l_{1}-4}+\frac{1}{2}k^2{\bar{T}}^{4(l_{1}-1)}+
\alpha^2{\bar{T}}^{\frac{4(l_{1}-1)}{3}}e^{-3M\bar{T}^{l_{1}}}
\]
\begin{equation}
\label{eq42} + 24\pi\xi\left(n{\bar{T}}^{l_{1}-2}+\frac{1}{2}k{\bar{T}}^{2(l_{1}-1)}\right)
\end{equation}
\begin{equation}
\label{eq43} 8\pi\rho=3n^2{\bar{T}}^{2(l_{1}-2)}+3nk{\bar{T}}^{3l_{1}-4}+\frac{1}{2}k^2{\bar{T}}^{4(l_{1}-1)}-
3\alpha^2{\bar{T}}^{\frac{4(l_{1}-1)}{3}}e^{-3M\bar{T}^{l_{1}}},
\end{equation}
\begin{equation}
\label{eq44} 8\pi\rho_{p}=n(3n-2){\bar{T}}^{2(l_{1}-2)}-4\alpha^2{\bar{T}}^{\frac{4(l_{1}-1)}{3}}e^{3M\bar{\bar{T}}^{l_{1}}} 
- 24\pi\xi\left(n{\bar{T}}^{l_{1}-2}+\frac{1}{2}k{\bar{T}}^{2(l_{1}-1)}\right) 
\end{equation}
\begin{figure}
\begin{center}
\includegraphics[width=4.0in]{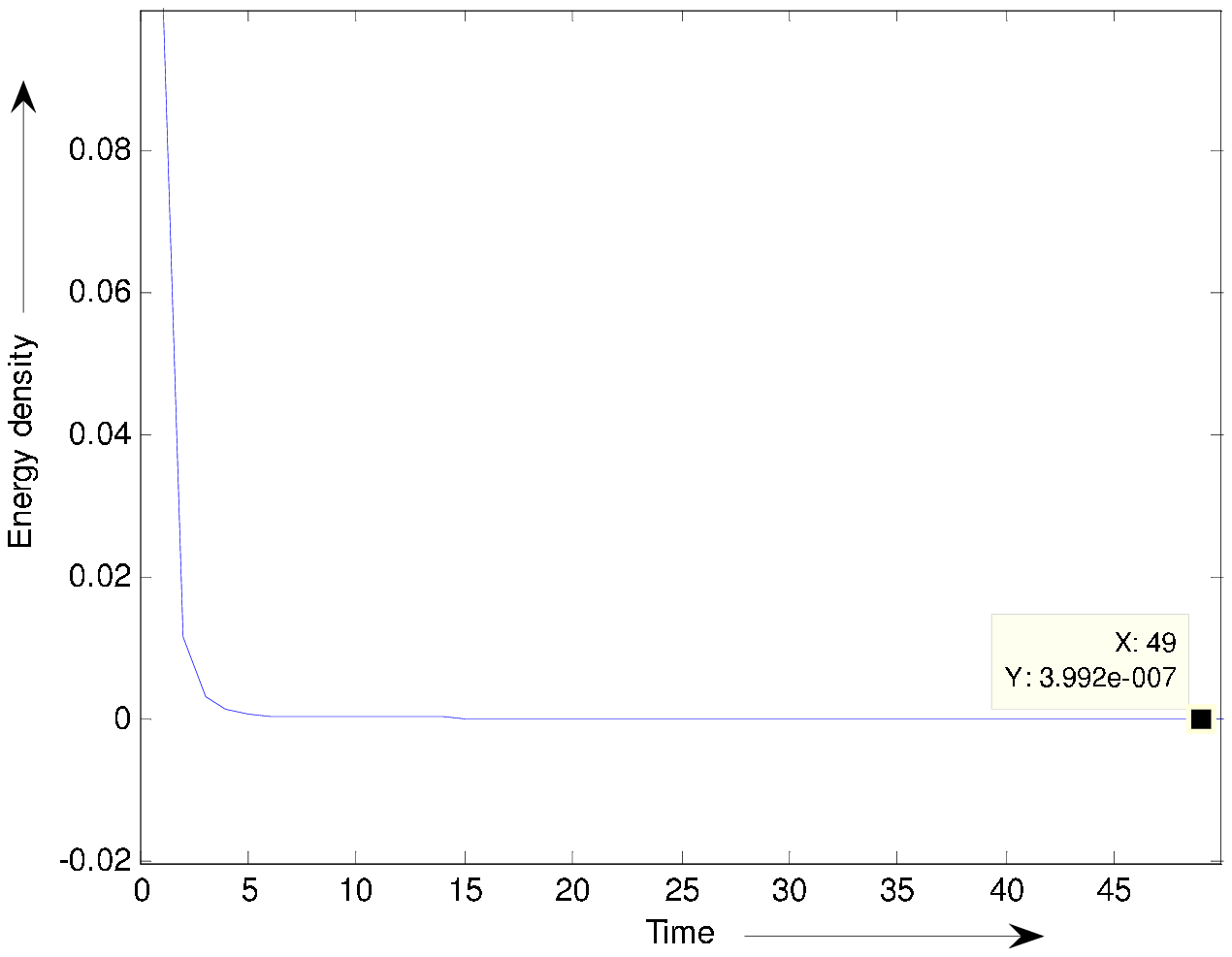} 
\caption{The plot of energy density $\rho$ vs. time $\bar{T}$ with parameters $ n = 0.5, k = 1, \alpha = 1 
 k_{1} = 1, l_{1} = 0.25 $}
\label{fg:anil22F2.eps}
\end{center}
\end{figure}
%%%%%%%%%%%%%%%%%%%%%%%%%%%%%%%%%%%%%%%%%%%%%%%%%%%%%%%%%%%%%%%%%%%%%%%%%%%%%%%%%%%%%%%%%%%%%%%%%%%%%%%
%%%%%%%%%%%%%%%%%%%%%%%%%%%%%%%%%%%% Subsubsection 3.2.1 %%%%%%%%%%%%%%%%%%%%%%%%%%%%%%%%%%%%%%%%%%%%%%%
\subsubsection{Case I:~ Solution for $\xi = \xi_{0}$}  
For $ w = 0 $, eqs.(\ref{eq27}) reduces $\xi=\xi_{0}$.Thus eqs.(\ref{eq42}) and eqs.(\ref{eq44}) reduces to
\[
 8\pi\lambda=2n{\bar{T}}^{2(l_{1}-2)}+3nk{\bar{T}}^{3l_{1}-4}+\frac{1}{2}k^2{\bar{T}}^{4(l_{1}-1)}+
\alpha^2{\bar{T}}^{\frac{4(l_{1}-1)}{3}}e^{-3M\bar{T}^{l_{1}}} + 
\]
\begin{equation}
\label{eq45} 24\pi\xi_{0}\left[n{\bar{T}}^{l_{1}-2}+\frac{1}{2}k{\bar{T}}^{2(l_{1}-1)}\right]
\end{equation} 
\begin{equation}
\label{eq46} 8\pi\rho_{p}=n(3n-2){\bar{T}}^{2(l_{1}-2)}-4\alpha^2{\bar{T}}^{\frac{4(l_{1}-1)}{3}}e^{3M\bar{\bar{T}}^{l_{1}}} 
- 24\pi\xi_{0}\left[n{\bar{T}}^{l_{1}-2}+\frac{1}{2}k{\bar{T}}^{2(l_{1}-1)}\right] 
\end{equation}
%%%%%%%%%%%%%%%%%%%%%%%%%%%%%%%%%%%%%%%%%%%%%%%%%%%%%%%%%%%%%%%%%%%%%%%%%%%%%%%%%%%%%%%%%%%%%%%%%%%%%%
%%%%%%%%%%%%%%%%%%%%% Subsubsection 3.2.2 %%%%%%%%%%%%%%%%%%%%%%%%%%%%%%%%%%%%%%%%%%%%%%%%%%%%%%%%%%%%
\subsubsection{Case II:~ Solution for $\xi = \xi_{0}\rho$}
For $w=1$, eqs.(\ref{eq27}) reduces $\xi=\xi_{0}\rho$, with use of eqs.(\ref{eq43}), eqs.(\ref{eq42}) and 
eqs.(\ref{eq44}) reduces to
\[
 8\pi\lambda=2n{\bar{T}}^{2(l_{1}-2)}+3nk{\bar{T}}^{3l_{1}-4}+\frac{1}{2}k^2{\bar{T}}^{4(l_{1}-1)}+
\alpha^2{\bar{T}}^{\frac{4(l_{1}-1)}{3}}e^{-3M\bar{T}^{l_{1}}}+ 
\]
\[
 3\xi_{0}\left[3n^2{\bar{T}}^{2(l_{1}-2)}+3nk{\bar{T}}^{3l_{1}-4}+\frac{1}{2}k^2{\bar{T}}^{4(l_{1}-1)}-
3\alpha^2{\bar{T}}^{\frac{4(l_{1}-1)}{3}}e^{-3M\bar{T}^{l_{1}}}\right]
\]
\begin{equation}
\label{eq47} 
\left(n{\bar{T}}^{l_{1}-2}+\frac{1}{2}k{\bar{T}}^{2(l_{1}-1)}\right)
\end{equation}
\[
8\pi\rho_{p}=n(3n-2){\bar{T}}^{2(l_{1}-2)}-4\alpha^2{\bar{T}}^{\frac{4(l_{1}-1)}{3}}
e^{3M{\bar{T}}^{l_{1}}}- 3\xi_{0}\left(n{\bar{T}}^{l_{1}-2}+\frac{1}{2}k{\bar{T}}^{2(l_{1}-1)}\right)
\]
\begin{equation}
\label{eq48} \left[3n^2{\bar{T}}^{2(l_{1}-2)}+3nk{\bar{T}}^{3l_{1}-4}+\frac{1}{2}k^2{\bar{T}}^{4(l_{1}-1)}-
3\alpha^2{\bar{T}}^{\frac{4(l_{1}-1)}{3}}e^{-3M\bar{T}^{l_{1}}}\right] 
\end{equation}
From Eqs.(\ref{eq43}), we note that $\rho(t)$ is a decreasing function of time
and $\rho > 0$ for all times. This behaviour is clearly shown in Fig. 2. 
Also it is observed that bulk viscosity affect the string tension density $\lambda$ and particle density $\rho_{p}$. 
\begin{equation}
\label{eq49} \theta=3\left[n{\bar{T}}^{l_{1}-2}+\frac{1}{2}k{\bar{T}}^{2(l_{1}-1)}\right],
\end{equation}
\begin{equation}
\label{eq50} \sigma=\frac{1}{2}k\bar{T}^{2(l_{1}-1)}e^{-3M\bar{T}^{l_{1}}} 
\end{equation}
\begin{equation}
\label{eq51} V^{3}=\left(k_{3}\bar{T}^{2n}e^{{M}\bar{T}^{l_{1}}}\right)^{\frac{3}{2}}e^{2\alpha x}
\end{equation}
From equation (\ref{eq49}) and (\ref{eq50}) leads to
\begin{equation}
\label{eq52} \frac{\sigma}{\theta}=\frac{k}{6(n\bar{T}^{-l_{1}}+\frac{1}{2}k)}
\end{equation}
The energy condition $ \rho\geq0 $ and $ \rho_{p}\geq0 $ are satisfy for model (\ref{eq36}). The condition $ \rho\geq0 $ and
$ \rho_{p}\geq0 $ leads to\\
\begin{equation}
\label{eq53} e^{3M\bar{T}^{l_{1}}}\left[3n^2\bar{T}^{\frac{2l_{1}-8}{3}}+3nk\bar{T}^{\frac{5l_{1}-8}{3}}+
\frac{1}{2}k^2\bar{T}^{\frac{8l_{1}-8}{3}}\right]\geq3\alpha^2
\end{equation}
\begin{equation}
\label{eq54}{\bar{T}}^{\frac{(1-4l_{1})}{3}}e^{-3M\bar{\bar{T}}^{l_{1}}}\left[n(3n-2){\bar{T}}^{2(l_{1}-2)}
 - 24\pi\xi\left(n{\bar{T}}^{l_{1}-2}+\frac{1}{2}k{\bar{T}}^{2(l_{1}-1)}\right)\right] \geq4\alpha^2 
\end{equation}
respectively.\\
we observe that the string tension density $\lambda\geq0$, leads to
\[
 {\bar{T}}^{\frac{(1-4l_{1})}{3}}e^{3M\bar{T}^{l_{1}}}\times
\]
\begin{equation}
\label{eq55} \left[2n{\bar{T}}^{2(l_{1}-2)}+3nk{\bar{T}}^{3l_{1}-4}+
\frac{1}{2}k^2{\bar{T}}^{4(l_{1}-1)}+24\pi\xi\left(n{\bar{T}}^{l_{1}-2}+\frac{1}{2}k{\bar{T}}^{2(l_{1}-1)}\right)\right]
\geq-\alpha^2 
\end{equation}
For $l_{1}>2$, model (\ref{eq41}) is expanding and for $ l_{1}<2 $, model starts with big bang singularity.
Generally the model is expanding, shearing and approaches isotropy at late time. For $k=0$, 
the model becomes shear free. We observe that as $\bar{T}\rightarrow\infty$, $V^3\rightarrow\infty$ and $\rho\rightarrow0$
hence volume increases when $\bar{T}$ increases and the proper energy density of the cloud of string with particle 
attached to them is the decreasing function of time.\\ 
%%%%%%%%%%%%%%%%%%%%%%%%%%%%%%%%%%%%%%%%%%%%%%%%%%%%%%%%%%%%%%%%%%%%%%%%%%%%%%%%%%%%%%%%%%%%%%%%%%%%%%%%%%%%%%%%%%%%%%%%%%
%%%%%%%%%%%%%%%%%%%%%%%%%%%%%%%%Subsection 3.3%%%%%%%%%%%%%%%%%%%%%%%%%%%%%%%%%%%%%%%%%%%%%%%%%%%%%%%%%%%%%%%%%%%%%%%%%%%
\subsection{Case(iii): $ B={\tilde{t}}^n $~~(n is a real number)}
In this case equation (\ref{eq19}) gives and
\begin{equation}
\label{eq56} C=k_{3}{\tilde{t}^n}-2l{\tilde{t}}^{l_{1}}
\end{equation}
\begin{equation}
\label{eq57} A^2=k_{3}{\tilde{t}}^{2n}-2l{\tilde{t}}^{l_{1}+n}
\end{equation}
Hence the metric $(1)$ takes the new form
\begin{equation}
\label{eq58} ds^2=\left(k_{3}{\tilde{t}}^n-2l{\tilde{t}}^{l_{1}}\right)\left[dt^2-{\tilde{t}}^{n}dx^2\right]
-e^{2\alpha x}\left[{\tilde{t}}^{2n}dy^2+\left(k_{3}{\tilde{t}}^{n}-2l{\tilde{t}}^{l_{1}}\right)^{2}dz^2\right],
\end{equation}
In this case the physical parameters, i.e. the string tension density $(\lambda)$, the energy density $(\rho)$, 
the particle density $(\rho_{p})$ and the kinematical parameters, i.e. the scalar of expansion $(\theta)$,
shear scalar $(\sigma)$ and the proper volume $(V^3)$ for model (\ref{eq58}) are given by\\
\[
 8\pi\lambda=\left[-\frac{1}{2}l^{2}(11n^2-4n-4){\tilde{t}}^{-3n}+lk_{3}n(13n-10){\tilde{t}}^{l_{1}+n}
-2k_{3}^2n(n-1){\tilde{t}}^{2n-2}\right]\psi_{2}^{-3}
\]
\begin{equation}
\label{eq59} +\alpha^2{\tilde{t}}^{-n}\psi_{2}^{-1} + 24\pi\xi\left(\frac{1}{2}l(n-2){\tilde{t}}^{-\frac{3n}{2}}+
k_{3}n\tilde{t}^{n-1}\right)\psi_{2}^{-\frac{3}{2}},
\end{equation}
\[
 8\pi\rho=\left[-\frac{1}{2}l^{2}(11n^2-4n-4){\tilde{t}}^{-3n}-3lk_{3}n(2-n){\tilde{t}}^{l_{1}+n}
+3k_{3}^{2}n^2{\tilde{t}}^{2n-2}\right]\psi_{2}^{-3} -
\]
\begin{equation}
\label{eq60} 3\alpha^2{\tilde{t}}^{-n}\psi_{2}^{-1}
\end{equation}
\[
 8\pi\rho_{p}=\psi_{2}^{-3}\left[
  n(5n-2)k_{3}^2\tilde{t}^{2n-2}-lk_{3}n(12n-8)\tilde{t}^{l_{1}+n}\right]-4\alpha^2\tilde{t}^n\psi_{2}^{-1} -
 \]
\begin{equation}
\label{eq61}  24\pi\xi\left(\frac{1}{2}l(n-2){\tilde{t}}^{-\frac{3n}{2}}+
k_{3}n\tilde{t}^{n-1}\right)\psi_{2}^{-\frac{3}{2}}, 
\end{equation}
\begin{figure}
\begin{center}
\includegraphics[width=4.0in]{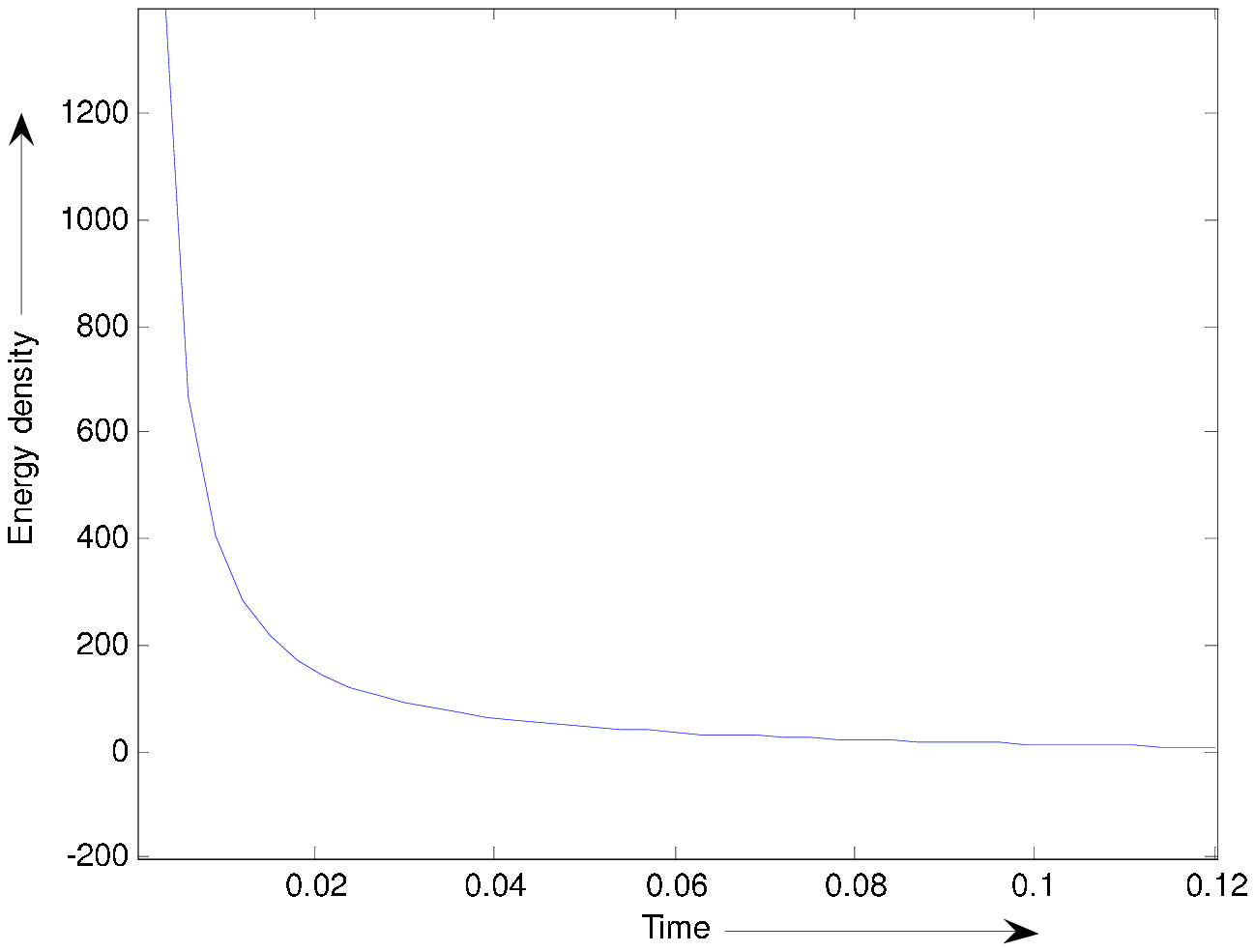} 
\caption{The plot of energy density $\rho$ vs. time $\bar{T}$ with parameters $ n = 0.5, k = 1, \alpha = 1 
 k_{1} = 1, l_{1} = 0.25 $}
\label{fg:anil22F3.eps}
\end{center}
\end{figure}
%%%%%%%%%%%%%%%%%%%%%%%%%%%%%%%%%%%%%%%%%%%%%%%%%%%%%%%%%%%%%%%%%%%%%%%%%%%%%%%%%%%%%%%%%%%%%%%%%%%%%%%
%%%%%%%%%%%%%%%%%%%%%%%%%%%%%%%%%%%% Subsubsection 3.2.1 %%%%%%%%%%%%%%%%%%%%%%%%%%%%%%%%%%%%%%%%%%%%%%%
\subsubsection{Case I:~ Solution for $\xi = \xi_{0}$}  
For $ w = 0 $, eqs.(\ref{eq27}) reduces $\xi=\xi_{0}$.Thus eqs.(\ref{eq59}) and eqs.(\ref{eq61}) reduces to
\[
 8\pi\lambda=\left[-\frac{1}{2}l^{2}(11n^2-4n-4){\tilde{t}}^{-3n}+lk_{3}n(13n-10){\tilde{t}}^{l_{1}+n}
-2k_{3}^2n(n-1){\tilde{t}}^{2n-2}\right]\psi_{2}^{-3}
\]
\begin{equation}
\label{eq62} +\alpha^2{\tilde{t}}^{-n}\psi_{2}^{-1} + 24\pi\xi_{0}\left(\frac{1}{2}l(n-2){\tilde{t}}^{-\frac{3n}{2}}+
k_{3}n\tilde{t}^{n-1}\right)\psi_{2}^{-\frac{3}{2}},
\end{equation}
\[
 8\pi\rho_{p}=\psi_{2}^{-3}\left[
  n(5n-2)k_{3}^2\tilde{t}^{2n-2}-lk_{3}n(12n-8)\tilde{t}^{l_{1}+n}\right]-4\alpha^2\tilde{t}^n\psi_{2}^{-1} -
 \]
\begin{equation}
\label{eq63} 24\pi\xi_{0}\left[\frac{1}{2}l(n-2){\tilde{t}}^{-\frac{3n}{2}}+
k_{3}n\tilde{t}^{n-1}\right]\psi_{2}^{-\frac{3}{2}}, 
\end{equation}
%%%%%%%%%%%%%%%%%%%%%%%%%%%%%%%%%%%%%%%%%%%%%%%%%%%%%%%%%%%%%%%%%%%%%%%%%%%%%%%%%%%%%%%%%%%%%%%%%%%%%%
%%%%%%%%%%%%%%%%%%%%% Subsubsection 3.2.2 %%%%%%%%%%%%%%%%%%%%%%%%%%%%%%%%%%%%%%%%%%%%%%%%%%%%%%%%%%%%
\subsubsection{Case II:~ Solution for $\xi = \xi_{0}\rho$}
For $w=1$, eqs.(\ref{eq27}) reduces $\xi=\xi_{0}\rho$, with use of eqs.(\ref{eq59}), eqs.(\ref{eq61}) and 
eqs.(\ref{eq44}) reduces to
\[
 8\pi\lambda=\left[-\frac{1}{2}l^{2}(11n^2-4n-4){\tilde{t}}^{-3n}+lk_{3}n(13n-10){\tilde{t}}^{l_{1}+n}
-2k_{3}^2n(n-1){\tilde{t}}^{2n-2}\right]\psi_{2}^{-3}
\]
\[
+\alpha^2{\tilde{t}}^{-n}\psi_{2}^{-1} - 9\xi_{0}\left(\frac{1}{2}l(n-2){\tilde{t}}^{-\frac{3n}{2}}+
k_{3}n\tilde{t}^{n-1}\right)\psi_{2}^{-\frac{5}{2}} +
\]
\[
 3\xi_{0}\left[-\frac{1}{2}l^{2}(11n^2-4n-4){\tilde{t}}^{-3n}
-3lk_{3}n(2-n){\tilde{t}}^{l_{1}+n}+3k_{3}^{2}n^2{\tilde{t}}^{2n-2}\right]\psi_{2}^{-3}\times
\]
\begin{equation}
\label{64} \left(\frac{1}{2}l(n-2){\tilde{t}}^{-\frac{3n}{2}}+
k_{3}n\tilde{t}^{n-1}\right)\psi_{2}^{-\frac{3}{2}},
\end{equation}
\[
 8\pi\rho_{p}=\psi_{2}^{-3}\left[n(5n-2)k_{3}^2\tilde{t}^{2n-2}-lk_{3}n(12n-8)\tilde{t}^{l_{1}+n}\right]-
4\alpha^2\tilde{t}^n\psi_{2}^{-1}+
 \]
\[
  9\xi_{0}\left(\frac{1}{2}l(n-2){\tilde{t}}^{-\frac{3n}{2}}+
k_{3}n\tilde{t}^{n-1}\right)\psi_{2}^{-\frac{5}{2}} - 3\xi_{0}\left(\frac{1}{2}l(n-2){\tilde{t}}^{-\frac{3n}{2}}+
k_{3}n\tilde{t}^{n-1}\right)\psi_{2}^{-\frac{3}{2}}\times  
\]
\begin{equation}
\label{65} \left[-\frac{1}{2}l^{2}(11n^2-4n-4){\tilde{t}}^{-3n}
-3lk_{3}n(2-n){\tilde{t}}^{l_{1}+n}+3k_{3}^{2}n^2{\tilde{t}}^{2n-2}\right]\psi_{2}^{-3}, 
\end{equation}
From Eqs.(\ref{eq60}), we note that $\rho(t)$ is a decreasing function of time
 This behaviour is clearly shown in Fig. 3. 
Also it is observed that bulk viscosity affect the string tension density $\lambda$ and particle density $\rho_{p}$. 
\begin{equation}
\label{eq66} \theta=3\left[\frac{1}{2}l(n-2){\tilde{t}}^{-\frac{3n}{2}}+k_{3}n\tilde{t}^{n-1}\right]
\psi_{2}^{-\frac{3}{2}},
\end{equation}
\begin{equation}
\label{eq67} \sigma=\frac{1}{2}k{\tilde{t}}^{\frac{-3n}{2}}\psi_{2}^{-\frac{3}{2}},
\end{equation}
\begin{equation}
\label{eq68} V^3=\left(k_{3}{\tilde{t}}^{2n}-2l{\tilde{t}}^{l_{1}+n}\right)^{\frac{3}{2}}e^{2\alpha x}~~~~~~~~~~~~~~~~~~~~~~~~~~~~
\end{equation}
Equation (\ref{eq66}) and (\ref{eq54}) leads to
\begin{equation}
\label{eq69} \frac{\sigma}{\theta}=\frac{k}{6}\left[k_{3}n{\tilde{t}}^{-(l_{1}+n)}+\frac{l(n-2)}{2}\right]^{-1}
\end{equation}
Where $ \psi_{2} = \left(k_{3}{\tilde{t}}^n-2l{\tilde{t}}^{l_{1}}\right) $.\\
The energy condition $\rho\geq0$ and $\rho_{p}\geq0$ are satisfy for model (\ref{eq58}). The condition $\rho\geq0$ and
$ \rho_{p}\geq0 $ leads to\\
\begin{equation}
\label{eq70} \left[-\frac{1}{2}l^2\left(11n^2-4n-4\right)\tilde{t}^{-2n}-3lk_{3}n(2-n)\tilde{t}^{l_{1}+2n}+3k_{3}^2n^2
\tilde{t}^{(3n-2)}\right]\psi_{2}^{-2}\geq3\alpha^2
\end{equation}
\[
 \left(k_{3}\tilde{t}^n-2l\tilde{t}^{l_{1}}\right)^{-2}\left[n(5n-2)k_{3}^2\tilde{t}^{n-2}-lk_{3}n(12n-8)\tilde{t}^
l_{1}\right]-24\pi\xi\times
\]
\begin{equation}
\label{eq71} \left[\frac{1}{2}l(n-2){\tilde{t}}^{-\frac{n}{2}}+k_{3}n\tilde{t}^{-1}\right]
\psi_{2}^{-\frac{1}{2}}\geq4\alpha^2
\end{equation}
respectively.\\
We observe that the string tension density $\lambda\geq0$, leads to
\vspace{2mm}
\[
 \left[-\frac{1}{2}l^2\left(11n^2-4n-4\right)\tilde{t}^{-2n}+lk_{3}(13n-10)\tilde{t}^{l_{1}+2n}-2k_{3}^2n(n-1)
 \tilde{t}^{3n-2}\right]\times 
\]
\begin{equation}
\label{eq72} \psi_{2}^{-2}+24\pi\xi\left[\frac{1}{2}l(n-2){\tilde{t}}^{-\frac{n}{2}}+k_{3}n\tilde{t}^{-1}\right]
\psi_{2}^{-\frac{1}{2}}\geq-\alpha^2
\end{equation}
Thus we see that the model (\ref{eq58}) is generally expanding, shearing and approaches to isotropy at late time.
For $k=0$,the model becomes shear free. We observe that as $\tilde{t}\rightarrow\infty$, $V^3\rightarrow\infty$ and $\rho\rightarrow0$
hence volume increases when $\tilde{t}$ increases and the proper energy density of the cloud of string with particle 
attached to them is the decreasing function of time.\\ 

%%%%%%%%%%%%%%%%%%%%%%%%%%%%%%%%%%%%%%%%%%%%%%%%%%%%%%%%%%%%%%%%%%%%%%%%%%%%%%%%%%%%%%%%%%%%%%%%%%%%%%%%%%%%%%%%%%%%%%
%%%%%%%%%%%%%%%%%%%%%%%%%%%%%%%%%%%%%%%%%%%%Subsection3.4%%%%%%%%%%%%%%%%%%%%%%%%%%%%%%%%%%%%%%%%%%%%%%%%%%%%%%%%%%%%%
\subsection{Case(iv): $ B=\tau^n $~~ (n is any real number)}
In this case equation (\ref{eq20}) gives
\begin{equation}
\label{eq73} C=k_{4}\tau^ne^{\left(\frac{k}{l_{1}}\tau^{l_{1}}\right)}
\end{equation}
and then from equation (\ref{eq13}) we obtain
\begin{equation}
\label{eq74} A^2=k_{4}\tau^{2n}e^{\left(\frac{k}{l_{1}}\tau^{l_{1}}\right)}
\end{equation}
Hence the metric (\ref{eq1}) reduces to
\begin{equation}
\label{eq75} ds^2=\tau^{2n}e^{\left(\frac{k}{l_{1}}\tau^{l_{1}}\right)}\left[\tau^ne^{\left(\frac{2k}{l_{1}}\tau^{l_{1}}\right)}d\tau^2
-dx^2\right]-e^{2\alpha x}\left[dy^2+e^{\left(\frac{2k}{l_{1}}\tau^{l_{1}}\right)}dz^2\right],
\end{equation}
where without any loss of generality the constant $k_{4}$ is taken equal to $1$. Here we see that the metric function
A, B, and C are exponential type function as that of in  case $(ii)$. Thus we conclude that physical and geometrical 
properties of the model are similar to model (\ref{eq41}).\\
%%%%%%%%%%%%%%%%%%%%%%%%%%%%%%%%%%%%%%%%%%%%%%%%%%%%%%%%%%%%%%%%%%%%%%%%%%%%%%%%%%%%%%%%%%%%%%%%%%%%%%%%%%%%%%%%%%%%%%%
%%%%%%%%%%%%%%%%%%%%%%%%%%%%%%%%%%%%%%%%%%%%%%%%Conclusion%%%%%%%%%%%%%%%%%%%%%%%%%%%%%%%%%%%%%%%%%%%%%%%%%%%%%%%%%%%%%
\section{Conclusion}
If we choose $\alpha=0$, metric $(1)$ becomes Bianchi type I metric, studied by several authors in different context.
In this paper, we have applied the generation technique followed by Camci et al \cite{ref46} and found string
cosmological models with bulk viscosity. It is shown that the Einstein's field equation are solvable for an arbitrary cosmic scale function.
Starting from a particular cosmic function, new classes of spatially homogeneous and anisotropic cosmological models have been 
investigated for which the string fluid are acceleration and rotation free but they do have expansion and shear.
It is also observed that the physical and geometrical behavior of models in all cases are similar. Generally the model 
are expanding, shearing and non rotating. All the models are isotropized at late time. Also we found that in all 
cases energy density is decreasing function of time thus bulk viscosity decreases with time as it is assumed to be 
simple power function of energy density.

In case $(i)$, for $n\geq1$, model (\ref{eq23}) start expanding with big bang singularity 
and for $n\leq0$, model (\ref{eq23}) preserve expanding nature as $T\rightarrow0$,~$\theta\rightarrow0$ and 
$T\rightarrow\infty$,~$\theta\rightarrow\infty$. It is also observed that for $k=0$, shear scalar $(\sigma)$ vanishes 
and model becomes isotropic.

In case $(ii)$, we observed that for $ l_{1}<2$, model (\ref{eq41}) started with big bang singularity
and expand through out the evolution of universe. For $l_{1}>2$, model (\ref{eq41}) also preserve the 
expanding nature as $\bar{T}\rightarrow0$,~$\theta\rightarrow0$ and $\bar{T}\rightarrow\infty$,~
$\theta\rightarrow\infty$. From equation (\ref{eq50}), it is clear that when $k=0$, the shear scalar $(\sigma)$ vanishes and
model becomes isotropy. In case $(iii)$, it is observed that model (\ref{eq58}) started with big bang singularity 
and expanding through the evolution of universe. From equation (\ref{eq67}), it is clear that for $k=0$, 
the shear scalar $(\sigma)$ vanishes and model isotropizes. In case $(iv)$, it is observed that the properties 
of the metric (\ref{eq75}) are the same as that of the solution (\ref{eq41}), i.e the case $(ii)$.

The effect of bulk viscosity is to produce a change in perfect fluid and hence
exhibit essential influence on the character of the solution. In Section $3$, we
have shown regular well behaviour of energy density and the expansion of the universe with time parameter.
We also observe that Murphy's conclusion \cite{ref41} about the
absence of a big bang type singularity in the infinite past in models with bulk
viscous fluid, in general, is not true. The results obtained by Myung and Cho
\cite{ref47} also show that, it is, in general, not valid, since for some cases big bang
singularity occurs in finite past. For $\xi(t) = 0$ i. e. in absence of bulk viscosity, we get the solutions 
presented in our earlier work \cite{ref48}. 
 
%\newline
%\nonumsection{References}

\end{document}